\documentclass[twocolumn,showpacs,preprintnumbers,amsmath,amssymb]{revtex4}

\usepackage{graphicx}
\usepackage{dcolumn}
\usepackage{bm}

\begin{document}


\title{Dynamics and nonequilibrium states in
the \\ Hamiltonian mean-field model:  A closer look}

\author{Dami\'an H. Zanette}
\email{zanette@cab.cnea.gov.ar}
\affiliation{Consejo Nacional de Investigaciones Cient\'{\i}ficas  y
T\'ecnicas, Centro At\'omico Bariloche and Instituto Balseiro, 8400
San Carlos de Bariloche, R\'{\i}o Negro, Argentina.
}

\author{Marcelo A. Montemurro}
\email{mmontemu@famaf.unc.edu.ar}
\affiliation{Facultad de Matem\'atica, Astronom\'{\i}a y F\'{\i}sica,
Universidad Nacional de C\'ordoba, Ciudad Universitaria, 5000 C\'ordoba,
Argentina.}

\date{\today}

\begin{abstract}
We critically revisit the evidence for the existence of quasistationary
states in the globally coupled XY (or Hamiltonian mean-field) model. A
slow-relaxation regime at long times is clearly revealed by numerical
realizations of the model, but no traces of quasistationarity are found
during the earlier stages of the evolution. We point out the nonergodic
properties of this system in the short-time range, which makes a
standard statistical description unsuitable. New aspects of the evolution
during the nonergodic regime, and of the energy distribution function in
the final approach to equilibrium, are disclosed.
\end{abstract}

\maketitle

The globally coupled version of the XY model, given by
the Hamiltonian
\begin{equation} \label{H}
{\cal H} = K+V= \frac{1}{2}\sum_{i=1}^N p_i^2+
\frac{1}{2N} \sum_{i,j=1}^N [1-\cos (\theta_i-\theta_j)],
\end{equation}
has recently attracted considerable attention
\cite{RuffoPRE95,RuffoPRL98,RuffoPHD99,RuffoPRL99,TamPRL00,TsPRE01,TsPHA02,RuffoCSF02,mta}.
This model, usually referred to as the Hamiltonian mean-field (HMF) model,
describes an ensemble of $N$ rotators characterized by their angles
$\theta_i$ and conjugate momenta $p_i$. In contrast to the usual XY
model, interactions affect all pairs of rotators in the ensemble. The
HMF model has been studied in
connection with the emergence of collective self-organized states
\cite{RuffoPRE95},  dynamical behavior near phase transitions
\cite{RuffoPRL98}, and anomalous diffusion in phase space
\cite{RuffoPRL99}, among other dynamical and thermodynamical aspects.

The canonical thermodynamical description of the HMF model can be
completely carried out \cite{RuffoPRE95,RuffoPHD99}. In this
description, it is useful to introduce the  ``magnetization''
${\bf M}=N^{-1} \sum_i (\cos \theta_i, \sin \theta_i) \equiv (M_x,M_y)
\equiv M \exp({\rm i}\Phi)$ and the average energy per particle $U=E/N$,
with $E\equiv {\cal H}$.
The canonical approach predicts that the values of the magnetization
and the average energy at equilibrium, $M_{\rm eq}$ and $U_{\rm eq}$,  are
related by
\begin{equation} \label{U}
U_{\rm eq}=\frac{T}{2}+\frac{1}{2}(1-M_{\rm eq}^2),
\end{equation}
where $T$ is the canonical temperature. The system undergoes a
ferromagnetic-like, second-order phase transition at $T_c=1/2$,
from a state with $M_{\rm eq} \neq 0$ at low temperatures to a state
of vanishing magnetization at high temperatures. The energy at
the transition is $U_c = 3/4$. These predictions are mostly confirmed
by the numerical solution of the HMF equations of motion at fixed
total energy, i.e. in a microcanonical scenario. In this situation
the equilibrium temperature is defined through the kinetic energy
\begin{equation} \label{Teq}
T_{\rm eq}=\frac{2}{N} K_{\rm eq},
\end{equation}
and $U=T_{\rm eq}/2+(1-M_{\rm eq}^2)/2$ [cf. Eq. (\ref{U})].

Deviations between the canonical predictions and the
microcanonical (numerical) results were however reported in the
region of energies just below the critical point, $0.5<U<U_c$,
where numerical realizations reveal extremely slow relaxation
toward the asymptotic state \cite{RuffoPRE95,RuffoPRL98}. Such
deviations, observed when the ensemble of rotators is prepared in
specially chosen initial conditions---namely, the so called
water-bag initial conditions---have been associated with the
existence of long-lived states where, in the thermodynamical
limit $N\to \infty$, the system would spend asymptotically long
times. In that limit, canonical equilibrium would never be
reached. It has been suggested that the HMF long-lived
states---which would replace canonical equilibrium in the
thermodynamical limit---would be well described by the
equilibrium distributions predicted by Tsallis's  generalization
of thermodynamics \cite{TsPRE01,TsPHA02,TsPHA01}. This
generalization, in  fact, relaxes the assumption of extensivity
of the Boltzmann-Gibbs formulation, and is therefore presumably
expected to describe the equilibrium statistics of systems with
long-range interactions \cite{TsPHA95}, as in the Hamiltonian
(\ref{H}).


Water-bag (WB) initial conditions fix $\theta_i=0$ for all $i$,
and the momenta $p_i$ are chosen at random from a uniform distribution
in such a way that $\sum_i p_i=0$ and $K=\sum_i p_i^2/2 =NU$.
For these initial conditions, in fact, the potential energy
vanishes. For a typical numerical realization at energies just below the
critical point $U_c$, the dynamics of the HMF model can be summarized as
follows. In the first stage, whose duration is essentially independent of
the system size $N$ and equals a few time units \cite{TsPHA02}, there is
a rapid broadening of the distribution in $\theta$. This implies an abrupt
growth of the potential energy from its initial value $V(t=0)=0$, to a
value  close to that predicted at canonical equilibrium, $V_{\rm eq}=N(1-
M_{\rm eq}^2)/2$. The kinetic energy drops accordingly, from $K(t=0)=NU$ to
a value close to $K_{\rm eq}=NT_{\rm eq}/2$, and the initial unbalance
between $K$ and $V$ is drastically reduced. As explained below, the details
of the evolution in this stage of kinetic  relaxation depend strongly on
the specific realization of the WB initial condition. Typically, however,
the potential energy exceeds $V_{\rm eq}$ and attains a maximum, while
$K$ reaches a minimum. From then on, the system enters a stage of slow
relaxation toward canonical equilibrium.

\begin{figure}[h]
\begin{center}
\resizebox{\columnwidth}{!}{\includegraphics{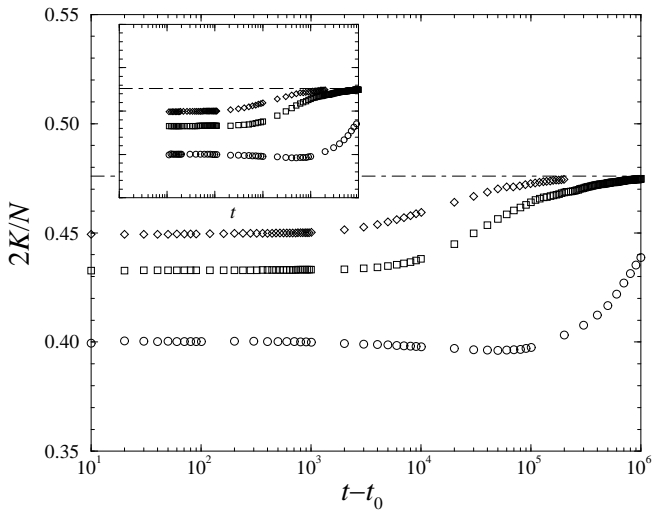}}
\end{center}
\caption{Nonequilibrium temperature $T(t)=2K(t)/N$ of the HMF model for
$U=0.69$ and $N=500$ (diamonds), $10^3$ (squares), and $10^4$ (circles),
from Fig.~1(b) of Ref.~\onlinecite{TsPRE01}. Note that data are plotted
against $t-t_0$, with  $t_0=100$. The horizontal line indicates the canonical
equilibrium temperature $T_{\rm eq}=0.476$. The inset shows the same data
against $t$. The scales of the axis in the inset are the same
as in the main plot.}
\label{f1}
\end{figure}

The slow-relaxation regime of the HMF model was reported in the first
systematic study of this system \cite{RuffoPRE95}. In
Ref.~\onlinecite{RuffoPHD99} it was quantitatively
characterized by analyzing the relaxation of the nonequilibrium entropy
$S(t)$ toward its equilibrium value $S_{\rm eq}$ as a function of $N$. The
difference  $|S(t)-S_{\rm eq}|$ was shown to approach an asymptotic value
of order  $N^{-1/2}$ within a relaxation time $\tau_r \sim N$. The same
scaling has recently been reported for the relaxation time of the
nonequilibrium temperature, $T(t)=2K(t)/N$ \cite{RuffoCSF02}. The
states visited by the system during the slow-relaxation regime have
been named quasistationary states, presumably by analogy with the
quasiequilibria reported for one-dimensional gravitational systems
\cite{grav1,grav2} and similar long-lived states in many-body systems
with long-range interactions \cite{Thirring}. The name
suggests that the time scales associated with the evolution of such
states are much longer than those of any other regime during the whole
evolution. However, the evolution of $S(t)$ toward $S_{\rm eq}$ reveals
no dynamical regime other than slow relaxation \cite{RuffoPHD99,init}.

Latora, Rapisarda and Tsallis (LRT), on the other hand, have
claimed to demonstrate the entity of quasitationary states (QSS)
from the analysis of the evolution of the nonequilibrium
temperature $T(t)$ \cite{TsPRE01}. For clarity in the discussion,
we reproduce in the main plot of Fig.~\ref{f1} their figure 1(b)
of Ref.~\onlinecite{TsPRE01}, where $T(t)$ is plotted against
time for several values of $N$, as obtained from numerical
integration of the HMF equations of motion for $U=0.69$
\cite{LA}. This value of the energy corresponds to the largest
observed deviation of numerical results with respect to the
canonical equilibrium prediction \cite{RuffoPRE95,RuffoPHD99}. In
the plot, the initial stage of kinetic relaxation is
disregarded.  For short times, we observe a wide plateau where
$T(t)$ is apparently constant, but substantially differs from its
equilibrium value. At long times, $T(t)$ has reached  $T_{\rm
eq}$  and remains at equilibrium. These two plateaus are
connected by a crossover stage where $T(t)$ exhibits an inflexion
point and its variation is seemingly faster. Consequently, LRT
conclude that the slow-relaxation regime consists in turn of  a
first stage where the system is ``trapped'' in a QSS, and a
subsequent  crossover toward canonical equilibrium. Since the
position $\tau_C$ of the crossover shifts to larger times as $N$
grows, $\tau_C \sim N$, LRT deduce that in the thermodynamical
limit the QSS will be observed at all times, and that canonical
equilibrium will never be attained. As we argue in the following,
however, the presence of  the plateaus and, thus, the existence
of a QSS in these specific  realizations of the HMF model, is an
artifact of the peculiar way chosen to plot the data in
Fig.~\ref{f1}.

In the first place, in order to exclude the initial stage of kinetic
relaxation, LRT choose to plot $T(t)$ as a function of $t-t_0$, with
$t_0=100$, instead of simply cutting off the time axis at $t_0$. Such
procedure, which  would be innocuous in a linear time scale, has
far-reaching consequences in the logarithmic scale of Fig.~\ref{f1}.
In fact, the point $t=t_0$ becomes shifted to $-\infty$ in the
logarithmic variable, which considerably enhances the impression of
having a plateau at small $t$. In the inset of Fig.~\ref{f1}
we plot $T(t)$ against its original variable---also in logarithmic
scale---showing that the short-time plateau shrinks by one decade.
The existence of a crossover state joining the two plateaus is itself
an illusion created by the logarithmic scale, even in the original
variable $t$. Indeed, it can be easily shown that the graph of a
function with monotonic first derivative can display an (otherwise
inexistent) inflection point by the simple expedient of using
a logarithmic scale in the horizontal axis \cite{log}.

\begin{figure}[h]
\begin{center}
\resizebox{\columnwidth}{!}{\includegraphics{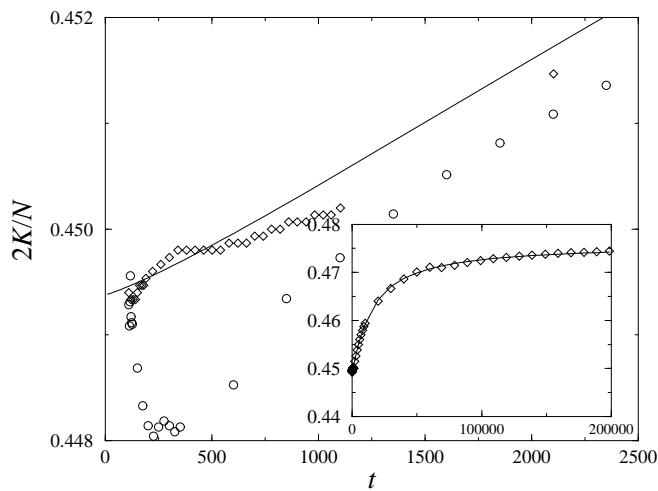}}
\end{center}
\caption{The same data as in Fig.~\ref{f1} for $N=500$ (diamonds),
with linear time scale. The inset shows the whole time range.
The curve is a Pad\'e-like approximation over the entire range.
Open circles are higher-precision results from Ref.~\onlinecite{TsPHA02}.}
\label{f2}
\end{figure}

\begin{figure}[h]
\begin{center}
\resizebox{\columnwidth}{!}{\includegraphics{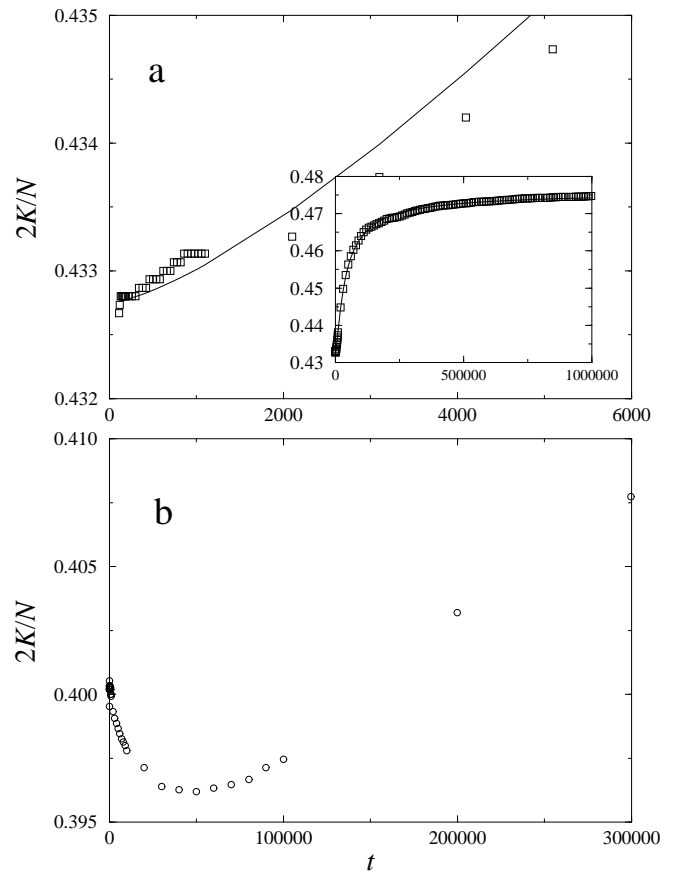}}
\end{center}
\caption{The same data as in Fig.~\ref{f1} for (a) $N=10^3$
and (b) $N=10^4$, with linear time scale. The inset in (a) shows
the whole time range, while the curve is a Pad\'e-like approximation
over the entire range.}
\label{f3}
\end{figure}

In order to avoid  these undesirable effects and to reveal the
true nature  of the evolution of $T(t)$ in the slow-relaxation
regime, we opt for using linear time scales. Figure \ref{f2}
shows, plotted with diamonds, the same data as in Fig.~\ref{f1}
for $N=500$ up to $t=2500$, while the inset shows  the whole
temporal range. We see that there are no traces of a short-time
plateau, except perhaps for a small interval around $t=500$. It
might however be argued that the vertical scale in this plot has
been exaggeratedly amplified. To decide over this point, we  have
used a Pad\'e-like approximation to fit the data over the whole
time range, shown as a curve in the inset.  The quality of the
fitting is quite acceptable. Its extension to the short-time
range, also shown as a curve in the main plot, proves that in
this range the data exhibit the same trend as for larger times.
The evolution in the whole range is therefore uniform, and we
find no arguments to attribute the condition of quasistationarity
to the short-time interval. As for the small interval around
$t=500$, we also include in Fig.~\ref{f2} a data set from
Ref.~\onlinecite{TsPHA02}, calculated with higher numerical
precision (open circles) \cite{LA2}. These data reveal a minimum
in $T(t)$ at $t\approx 300$ and confirm the absence of any
plateau. Figure \ref{f3} shows the data from Fig.~\ref{f1} for
$N=10^3$ and $10^4$ in linear time scale. Again, we find no
evidence of the existence of plateaus or QSS.

Let us mention that, more indirectly, the observation of
anomalous diffusion in the slow-relaxation stage of the HMF model
\cite{RuffoPRL99} confirms that no special regimes exist within
that stage. In fact, superdiffusion in the unfolded $\theta$-space,
$\theta \in (-\infty,+\infty)$, which is characterized by an anomalous
dispersion law $\langle\Delta \theta^2\rangle \sim t^{\alpha}$, is
recorded  in the whole slow-relaxation stage with a constant exponent
$\alpha\approx 1.4$. The transition to ordinary diffusion  ($\alpha=1$)
occurs at the same time as the system reaches canonical equilibrium.
The short-time plateau of Fig.~\ref{f1} and the crossover to
canonical equilibrium are not detected by the transport properties
of this system.

\begin{figure}[h]
\begin{center}
\resizebox{\columnwidth}{!}{\includegraphics{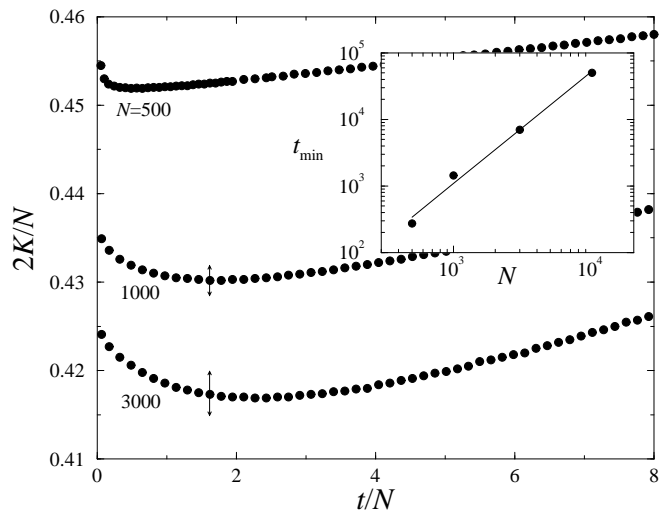}}
\end{center}
\caption{Nonequilibrium temperature, $T(t)=2K(t)/N$, for $U=0.69$
and three values of $N$, as a function of the rescaled time
$t/N$. The arrows indicate the typical dispersion of $T(t)$ after
averaging over a large set of realizations ($\sim 10^3$ to $\sim
10^4$, see text). For $N=500$ the dispersion is negligible in this
scale. The inset shows the time of the minimum in  $T(t)$ as a
function of $N$. The straight line is a power-law fit, with slope
$1.7$.}
\label{f4}
\end{figure}

The data shown in Figs.~\ref{f2} and \ref{f3} exhibit an
inconsistency regarding the behavior of the nonequilibrium
temperature in the first stage of the slow-relaxation regime. In
fact, while in the results for $N=500$  (circles) and for
$N=10^4$ we observe a well-defined minimum in $T(t)$,
low-precision data for $N=500$ (diamonds) and results for
$N=10^3$ do not display such minimum. To clarify this aspect of
the evolution, we have performed extensive numerical calculations
for the same value of the energy, $U=0.69$, maintaining the
relative error in the total energy conservation around $\Delta
E/E \sim 10^{-3}$, which makes it possible to average over many
realizations of the water-bag initial  conditions ($5\times 10^4$
for $N=500$ to $10^3$ for $N=3000$). Our results for the
nonequilibrium temperature $T(t)$ are shown in Fig.~\ref{f4}, as
a function of the rescaled time $t/N$. Each dot has been obtained
as an average in time, over $0.1N$ to $0.2N$ time units. The
arrows indicate, for $N=1000$ and $3000$, the typical dispersion
in the values of $T(t)$ after averaging over the whole set of
realizations. Fluctuations are indeed large, as expected near a
phase transition and due to the highly chaotic HMF dynamics
\cite{RuffoPRL98,RuffoPHD99}, which makes it difficult to obtain
reliable numerical results. In any case, our numerical
calculations consistently confirm the existence of a minimum in
the nonequilibrium temperature, and the subsequent steady growth
of $T(t)$. The inset shows the position of the minimum,
$t_{\min}$, as a function of $N$, including also the value for
$N=10^4$ taken from the data of Fig.~\ref{f3}(b). The straight
line corresponds to a power-law fitting, $t_{\min}\propto
N^{\nu}$, which yields $\nu=1.7 \pm 0.1$. Though the plot spans
less than two decades in $N$, the fact that $t_{\min}$ grows
faster than linearly with the system size is clear.

Meanwhile, no plateau is observed, at least, of the kind shown
in Fig.~\ref{f1}.  A plateau may of course be compulsively ascribed
to the zone around the minimum of $T(t)$ at $t_{\min}$, but it would be
far from coinciding in nature with the QSS reported by LRT. In
fact, although the minimum becomes broader for larger $N$, it also
shifts to longer times, which hardly insures that a stationary
(nonequilibrium) temperature would be observed in the limit
$N\to \infty$.

\begin{figure}[h]
\begin{center}
\resizebox{\columnwidth}{!}{\includegraphics{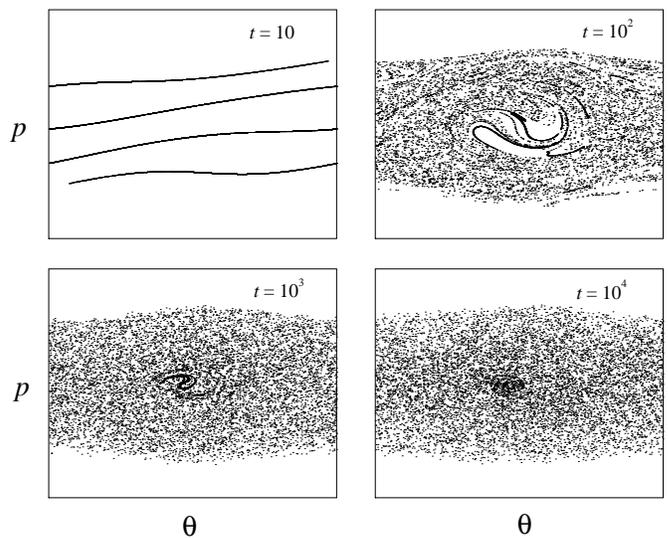}}
\end{center}
\caption{Snapshots in $\mu$-space at four different times of
a single realization of the HMF model with water-bag initial
conditions ($N=10^4$, $U=0.69$). In the four plots the horizontal
and vertical axes span the intervals $(-\pi,\pi)$ and $(-2,2)$,
respectively.}
\label{f5}
\end{figure}

The well-defined minimum of the nonequilibrium temperature $T(t)$
at $t_{\min}$ establishes a quite natural boundary between short-time and
long-time dynamical ranges. It is interesting to analyze in more detail
the short-time evolution, with emphasis in the angle and momentum distribution
of the ensemble in $\mu$-space, where at each time the system is represented
by a set of $N$ points in the $(\theta,p)$ plane. One-particle distributions
of both angles and momenta have received particular attention in previous work
\cite{RuffoPRE95,RuffoPHD99,TsPRE01,TsPHA02}, so that such representation is
useful for comparison.  Figure \ref{f5} shows four snapshots, at different
times, of the $\mu$-space distribution in a single (typical) realization of
the HMF model with $N=10^4$ rotators, at $U=0.69$. For clarity, the origin of
the angle variable, $\theta=0$, has been chosen to coincide in each plot
with the phase $\Phi$ of the total magnetization $\bf M$. In this way, the
center of the distribution coincides approximately with the center of the
plot. Take into account, however, that $\Phi$ performs, for short times,
diffusive-like motion \cite{RuffoPRE95}. The WB initial condition would
here correspond to a  uniform distribution along a vertical segment at
$\theta =0$, with $|p|<\sqrt{6U} \approx
2.03$. As advanced above, we observe a rapid broadening of the distribution
in the angle coordinate $\theta$, accompanied by a moderate collapse in $p$.
At $t=100$, much of the ensemble has become strongly mixed in $\mu$-space,
due to the highly chaotic nature of the dynamics for this value of $U$
\cite{RuffoPRE95,RuffoPHD99}. There is however a substantial fraction of
the ensemble, represented by the zones where points remain ordered, where
the strong correlations imposed by the WB initial condition persist.
This fraction is mainly situated near the center of the distribution, in
a region of relatively small momentum, but ordered arrays of points are
also seen in other zones of $\mu$-space. For these times, thus, the system
can be thought of as a mixture of two phases, namely, a strongly mixed,
``gaseous'' phase and a highly correlated, ``condensed'' phase.

\begin{figure}[h]
\begin{center}
\resizebox{.9\columnwidth}{!}{\includegraphics{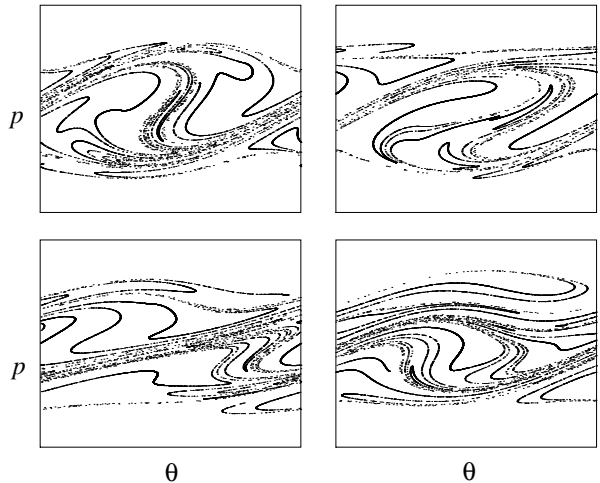}}
\end{center}
\caption{Snapshots in $\mu$-space at time $t=30$ of
four realizations of the HMF model with water-bag initial
conditions ($N=10^4$, $U=0.69$). Horizontal and vertical scales
are as in Fig.~\ref{f5}.}
\label{f6}
\end{figure}

The distribution of the condensed phase in $\mu$-space results to
strongly depend on the specific realization of the WB initial
condition. This sensibility to the initial conditions,
illustrated in Fig.~\ref{f6}, is again a consequence of the
chaotic dynamics that, as time elapses, amplifies the originally
small variations between different realizations. An important
byproduct of this property, combined with the persistence of the
initial correlations, is that the condensed phase is not ergodic,
in the sense that averages over realizations of the WB initial
condition yield a poor description of the statistical properties
over time of any single realization. The lack of ergodicity of
the condensed phase makes in turn the whole system nonergodic in
this short-time range. An illustration of this overall lack of
ergodicity is provided by the short-time evolution of the
nonequilibrium temperature. In any single realization, after its
initial drop, $T(t)$ displays chaotic oscillations of appreciable
amplitude. Due to the sensibility to initial conditions, however,
the phases of such oscillations differ between realizations, and
the oscillations disappear upon averaging. This observation
should warn of the averaging procedures used in previous work
\cite{TsPHA02} to study statistical properties such as the
nonequilibrium temperature and the momentum distribution function
in the initial regime of kinetic  relaxation. The warning applies
also to the average data of $T(t)$ shown in our Fig.~\ref{f4} at
the shortest times. Before the nonergodic condensate disappears,
in fact, a (standard) statistical approach to the HMF model
results to be of limited applicability.

As the evolution proceeds, particles from the condensate ``evaporate''
into the gaseous phase \cite{RuffoPRE95,RuffoPRL98}.
From Fig.~\ref{f5} we note
that, at $t=10^4$, only faint traces of the condensed phase remain in
the ensemble. We have studied the dynamics of this evaporation process
as follows. First of all, we have numbered the rotators in each
realization of the HMF model in such a way that, at the initial
time, $p_i(0)<p_j(0)$ for $i<j$. At  subsequent times, we have
defined the condensate as formed by the rotators $i$ such that the
distance to $i+1$ in $\mu$-space, $\delta_i = \sqrt{(\theta_{i+1}
-\theta_i)^2+(p_{i+1}-p_i)^2}$, is lower than a given threshold
$\delta$. We fix the threshold as the average distance between
contiguous pairs at $t=0$,  $\delta = 2\sqrt{6U}/N$.

\begin{figure}[h]
\begin{center}
\resizebox{\columnwidth}{!}{\includegraphics{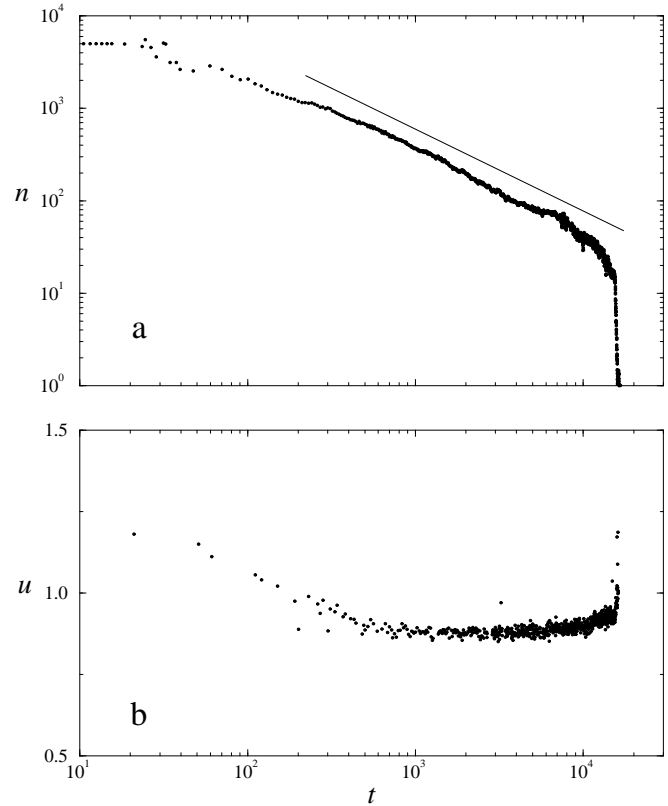}}
\end{center}
\caption{Evolution of (a) the size of the condensate phase and (b)
the average energy per particle in the condensate, for a single
realization of the HMF model with $N=10^4$ and $U=0.69$. Each dot
corresponds to a running average over $100$ time units. The
straight line in (a) has slope $-0.85$.}
\label{f7}
\end{figure}

The number $n$ of rotators in the condensed phase displays
considerable oscillations in time. According to the discussion in
the previous paragraph, we choose to measure $n$ in a single
(typical) realization for $N=10^4$ and to average over time
intervals. Figure \ref{f7}(a) shows the evolution of $n$ averaged
each $100$ time units. The condensate size decreases steadily,
displaying most of the time a power-law decay with a nontrivial
exponent, $n\sim t^{-0.85}$. This rather smooth decay is abruptly
interrupted by a sudden collapse of the condensate, at $t\approx
1.6 \times 10^4$. In Fig.~\ref{f7}(b), we show the evolution of
the average energy per particle in the condensate. We see that it
varies nonmonotonically with time, though its value is never far
from $u=1$. Other realizations for the same value of $N$ confirm
that these results are generic. Moreover, they are robust with
respect to the definition of the threshold $\delta$. For
different  values of $N$, realizations in the short-time range
show that the collapse of the condensed phase occurs at a time
preceding, but approximately equal to, $t_{\min}$. In other words,
when the minimum of the nonequilibrium temperature is attained,
the condensate has just evaporated. From then on, the system
enters the final stage of its relaxation to canonical equilibrium.

In the long-time range, when the ensemble of rotators has presumably
attained an ergodic state, its $\mu$-space configuration can be
described statistically, in terms of averages over realizations, by
means of a distribution function $f(\theta,p)$. Previous results on
the $\mu$-space distribution of the HMF model refer, in all cases, to
reduced distribution functions, $f_{\theta}(\theta)$ and $f_p(p)$. In
Ref.~\onlinecite{RuffoPHD99}, the momentum distribution function is
used to illustrate, for $U=0.69$, slow relaxation for long times.
For other values of $U$, far from the critical point, both $f_{\theta}
$ and $f_p$  were compared at moderately long times with the
prediction  of canonical  statistics, showing the relatively
fast attainment of canonical equilibrium for such energies. In
Refs.~\onlinecite{TsPRE01} and \onlinecite{TsPHA02}, in contrast,
deeper significance is attributed to nonequilibrium distribution
functions---always in connection with QSS. There, LRT suggest
that $f_p$ exhibits the fingerprints of nonextensive
thermodynamics, and support this claim showing that the momentum
distribution can be acceptably fitted, at low values of $p$, by
the power-law distribution derived from Tsallis's nonextensive
formulation \cite{TsPHA95}. Tsallis's formulation, however,
predicts the one-particle probability distribution for the energy,
and not for the momentum. Now, since the one-particle energy
\cite{RuffoPRE95}
\begin{equation} \label{h}
\epsilon = \frac{p^2}{2} +1-M_x \sin \theta -M_y \cos \theta
\end{equation}
depends both on $\theta$ and $p$, statistical
descriptions in terms of $\epsilon$ and $p$ are equivalent
only if the $\mu$-space distribution is independent of
$\theta$, i.e. if the energy distribution function $f_{\epsilon}
(\epsilon)$ can be unambiguously given in terms of
$f_p(p)$. This is not our case for $U<U_c$, unfortunately,
since the angle distribution is not uniform, as illustrated
by our Figs.~\ref{f5} and \ref{f6}, and by LRT's figure 3(a)
in Ref.~\onlinecite{TsPRE01}. Ignoring the angular dependence
of the one-particle distribution in $\mu$-space invalidates
the identification of $f_p$ as an equilibrium distribution
derived from a generalized thermodynamical formulation like
Tsallis's, which---just as ordinary thermodynamics---yields
canonical distributions as functions of the energy.

\begin{figure}[h]
\begin{center}
\resizebox{\columnwidth}{!}{\includegraphics{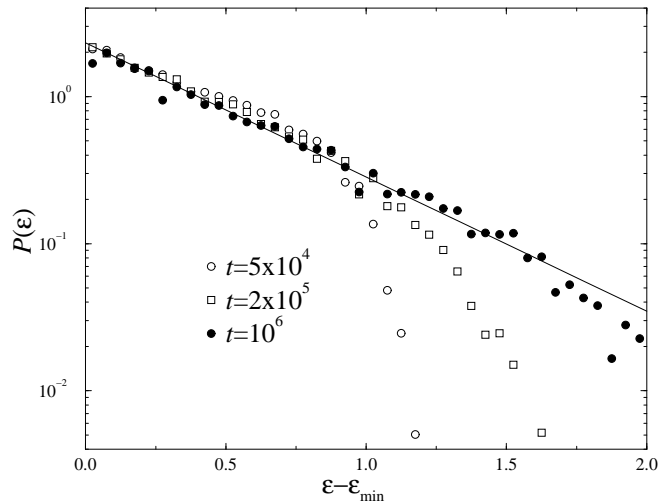}}
\end{center}
\caption{Probability distribution of energies at three times in
the long-time range, for a system of $N=10^4$ rotators at $U=0.69$,
as a function of $\epsilon -\epsilon_{\min}$. The straight line
stands for the exponential canonical prediction with temperature
$T=0.476$.}
\label{f8}
\end{figure}

We have evaluated $f_{\epsilon}$ by means of numerical
realizations of the HMF model for $N=10^4$. From the energy
distribution function we can in turn calculate the energy
probability density $P(\epsilon)$, since $f_{\epsilon}
(\epsilon)=\rho(\epsilon) P(\epsilon)$, where $\rho
(\epsilon)$ is the density of states. The HMF density of
states can be expressed in terms of elliptic functions, as
\begin{equation}
\rho (\epsilon) = \frac{2\sqrt{2}}{\pi \sqrt{\epsilon -1+M}}
K\left(\frac{2M}{\epsilon -1+M} \right)
\end{equation}
for $\epsilon>1+M$ and
\begin{equation}
\rho (\epsilon) = \frac{2\sqrt{2}}{\pi \sqrt{\epsilon
-1+M}} F\left(\frac{1}{2} \arccos \frac{1-\epsilon}{M},
\frac{2M}{\epsilon -1+M} \right)
\end{equation}
for $1-M<\epsilon<1+M$, where $K$ and $F$ are the complete
and  incomplete elliptic integrals of the first kind,
respectively  \cite{Ab}. The energy is bounded from below,
at $\epsilon_{\min} = 1-M$. Figure \ref{f8} shows the results for
$P(\epsilon)$ at three times, plotted as a function of $\epsilon
-\epsilon_{\min}$. The straight line represents the canonical
prediction, $P_{\rm eq}(\epsilon)\propto \exp(-\epsilon /T)$.
Observe that $t= 5\times 10^4$ corresponds to the minimum of the
nonequilibrium temperature, at the very beginning of the long-time
range. For this time we note a considerable deviation from canonical
equilibrium, with an abrupt cutoff at $\epsilon-\epsilon_{\min}
\approx 1$ preceded by an overpopulated interval. For very low energies,
however, the coincidence with the equilibrium distribution is already
quite good. For  $t=2 \times 10^5$ the cutoff persists but it is less
abrupt and has shifted to higher energies, while in the low-energy
range the distribution   is closer to canonical equilibrium. Finally,
for $t=10^6$, canonical equilibrium is well established in almost the
whole range of energies shown in the figure. Deviations of the
nonequilibrium temperature with respect to $T_{\rm eq}$ are therefore
to  be ascribed to the depleted tail at high energies. From these
results, we  conclude that in the long-time regime the equilibrium
energy distribution  of the HMF model with WB initial conditions
is built up from low energies, with a depleted high-energy tail
that recedes as time elapses.

We remark that LRT have assigned a distribution
function to states which are clearly in the short-time range.
In that range the system is still strongly nonergodic and, as
discussed above, such statistical description is therefore of very
limited significance. Their best fitting, for $N=10^5$ at $t=1200$,
corresponds in fact to a state where probably most of the ensemble
is still in the condensed phase. Moreover, they compare results
for different system sizes at the same time [see their figure 2(c)
in Ref.~\onlinecite{TsPRE01}], disregarding the scaling laws of
time scales with the size, which threatens the validity
of their evaluation of the fitting quality as a function of $N$
[their figure 2(d)]. Finally, we mention that the abrupt drop
of $f_p$ for large momenta---also seen at larger times in our
results for $P(\epsilon)$, Fig. \ref{f8}---eludes any description
based on Tsallis's thermodynamics. The artificial cutoff imposed
by LRT to their fitting of $f_p$, aimed at mimicking its abrupt
drop, can hardly find any consistent justification within such
formalism. We conclude that no evidence supports the association
of nonequilibrium HMF states with generalized thermodynamical
equilibria.

In summary,  we have reexamined the  slow-relaxation
dynamics of the Hamiltonian mean-field model, when the evolution
starts from water-bag initial conditions. First, we have revised
the evidence that would support the existence of quasistationary
states in the nonequilibrium dynamics of the model. In that
regard, we have shown that the plateaus in the temporal evolution
of the nonequilibrium temperature $T(t)$ \cite{TsPRE01,TsPHA02}
are a mere artifact of the use of a logarithmic scale for the time
axis in the plot of $T(t)$.  Moreover, a detailed  analysis of
the evolution of the  nonequilibrium temperature shows that the
dynamics can be decomposed into two  well-defined dynamical
regimes: a  short-time  range characterized  by the presence
of a condensed phase, in which the  system preserves strong
memory of the initial conditions, and a long-time range, in
which the condensed phase has evaporated and the system slowly
relaxes toward equilibrium. A  natural boundary  between these
two regimes is given by the minimum reached  by the nonequilibrium
temperature, which recedes toward larger times as the system
size grows. The numerical evaluation of the probability
density function of the one-particle energy in the long-time
range demonstrates that the system is slowly relaxing to the
Boltzmann distribution, and does not support any connection
with the generalized equilibrium distributions derived from
Tsallis's nonextensive formalism.

It must be stressed that neither previous work nor our present
results elucidate the nature of the HMF evolution in the
thermodynamical limit $N\to \infty$. The existence of different
dynamical regimes, with different scaling laws in terms of $N$,
makes an extrapolation from the available numerical results
extremely uncertain. The possibility of performing significant,
extensive numerical realizations of this model with current
computational facilities seems to be restricted to ensembles of,
at most, $N \sim 10^4$ elements. Numerical realizations for much
larger systems may be necessary in order to attain a meaningful
approach to the thermodynamical limit.

The origin of slow dynamics in the Hamiltonian mean-field model,
also revealed by the study of aging phenomena and strong memory
effects in the long-time range \cite{mta}, remains to be
explained as well. A plausible argument \cite{RuffoCSF02} may
perhaps arise from the study of the role of the state of zero
magnetization, which for $U<U_c$ is thermodynamically unstable,
and its connection with critical slowing down
\cite{RuffoPRL98}---a connection which has often been invoked but
not yet investigated. Revealing the basic mechanisms of slow
relaxation in the Hamiltonian mean-field model should definitely
clarify its relation to systems with some degree of structural
disorder or frustration, such as glasses, that also exhibit
slow-dynamics features. In turn, this would contribute to a
unified view on nonequilibrium phenomena in systems with many
degrees of freedom.

\begin{acknowledgments}
We thank S.~Ruffo for his motivating remarks on the HMF model,
and C.~Tsallis for comments on his own work. D.~H.~Z.~acknowledges
hospitality at the Abdus Salam International Centre for Theoretical
Physics (Trieste, Italy), where this work was partially carried out.
\end{acknowledgments}


\begin{thebibliography}{99}
\bibitem{grav2} Note that the definitions of the water-bag initial condition
respectively used for gravitational systems \cite{grav1} and for the HMF
model are  substantially  different. To our knowledge, the possible
consequences of this difference have not been discussed in the literature.
\end{thebibliography}
\end{document}